\documentclass[aps,preprint,groupedaddress,showpacs]{revtex4b1}

\begin{document}
\bibliographystyle{apsrev}

\title{Stopping of ion beam in a temperature anisotropic magnetized plasma}
\author{H. B. Nersisyan}
\thanks{Permanent address: Division of Theoretical Physics, 
Institute of Radiophysics and Electronics, 1 Alikhanian Brothers Str.,
Ashtarak -2, 378410, Armenia
\newline
E-mail: hrachya@irphe.am}


\author{M. Walter}
\author{G. Zwicknagel}
\affiliation{Institut f\"ur Theoretische Physik II, Universit\"at Erlangen, D - 91058 Erlangen, Germany}

\date{\today}

\begin{abstract}
Using the dielectric theory for a weakly coupled plasma we
investigate the stopping power of the ion in a temperature
anisotropic magnetized electron plasma. The analysis is based on the
assumption that the energy variation of the ion is much less than
its kinetic ener\-gy. The obtained general expression for stopping
power is analyzed for weak and strong magnetic fields (i.e.,
for the electron cyclotron frequency less than and greater than
the plasma frequency), and for low and high ion velocities. It is
found that the friction coefficient contains, in addition to the
usual velocity independent friction coefficient, an anomulous
term which diverges logarithmically as the projectile velocity
approaches zero. The physical origin of this anomulous term is the
coupling between the cyclotron motion of the electrons and
the long-wave length, low-frequency fluctuations produced by the
projectile ion.
\end{abstract}

\pacs{34.50.BW, 52.35.-g, 52.40.Mj}

\maketitle

\section{INTRODUCTION}

Energy loss of the ions in a plasma has been a topic of great interest
due to its considerable importance for the study of basic interactions
of the charged particles in real media. Recent applications are
electron cooling of heavy ion beams \cite{ahs,hp,inm} and energy transfer
for inertial confinement fusion (ICF) (see \cite{pro} for an overview).
Electron cooling is realized by mixing the ion beam periodically with
a cold electron beam of the same average velocity. The interaction
length is normally about a few meters and the electron beam is guided
by a magnetic field parallel to its direction of motion. The cooling
of the ion beam may then be viewed as an energy loss in the common
rest frame of both beams. Similar questions arise in
heavy-ion-induced ICF. There a frozen hydrogen pellet is heated and
compressed by stopping of ion beams in the surrounding converter. In
this case the electrons of the solid state converter are acting like a
plasma and absorb the incoming energy.

In the electron cooling process the velocity distribution of the electron beam is
highly anisotropic because of the acceleration from the cathode to the
cooling section. It can be described by a Maxwell distribution with two
different temperatures, a longitudinal $T_{\parallel}$ and a transversal
$T_{\bot}$ \cite{ahs,hp,inm}. Furthermore, an external, longitudinal magnetic field is
needed to guide the electrons from the cathode to and through the electron
cooler and to stabilize the anisotropic velocity distribution by suppressing
the transverse-longitudinal relaxation.

In the present paper we are interested in the influences of the magnetic
field and the temperature anisotropy on the ion beam stopping power.

Since the early 1960's several theoretical calculations of the stopping power
in a magnetized plasma have been presented \cite{iaa,nh,rmm,ggp,jgk,svb,hbn,
cd,cs,mw}. Stopping of a fast test
particle moving with velocity $V$ much higher than the electron thermal
velocity $v_{\rm th}$ was studied in Refs. \cite{iaa,nh,ggp}. Energy loss of a charged
particle moving with arbitrary velocity was studied in Ref. \cite{rmm}. The expression
obtained there for the Coulomb logarithm, $\Lambda =\ln (\lambda _D/\rho _{\perp })$
(where $\lambda _D$ is the Debye length and $\rho _{\perp }$ is the impact parameter
for scattering for an angle $\vartheta =\pi /2$), corresponds to the
classical description of collisions. In the quantum-mechanical case, the
Coulomb logarithm is $\Lambda =\ln (\lambda _D/\lambda _B)$, where $\lambda _B$ is
the de Broglie wavelength of plasma electrons \cite{el}.

In Ref. \cite{svb}, the expressions were derived describing the stopping power of
a charged particle in Maxwellian plasma placed in a classically strong (but
not quantizing) magnetic field ($\lambda _B\ll a_c\ll \lambda _D,$ where $%
a_c $ is the electron Larmor radius), under the conditions when scattering
must be described quantum mechanically. Calculations were carried out for
slow test particles whose velocities satisfy the conditions $%
(m/m_i)^{1/3}v_{\rm th}<V\ll v_{\rm th}$, where $m_i$ is the mass of the plasma
ions and $m$ is the electron mass.

In the recent paper \cite{hbn} the stopping power in the magnetized plasma has been
investigated for high-velocity light particles taking into account the Larmor
rotation of a test projectile in a magnetic field. It has been shown that the stopping
power can exhibit an oscillatory dependence on the magnetic field and that it is
much greater than in the case without magnetic field.

More attention has been paid on the stopping power in a strongly magnetized
plasma for ions which move along the magnetic field \cite{hbn,cd,cs}. Both
uncorrelated \cite{hbn,cs} and correlated \cite{cd} situations have been discussed.

These investigations have concentrated on the stopping power in temperature
isotropic plasma. Extensions to nonlinear effects
of ion stopping and temperature anisotropy have been done recently by
particle-in-cell (PIC) computer simulation \cite{mw}, where the case
$T_{\parallel} \ll T_{\bot}$ has been investigated which is interesting
for electron cooling process. Here, in the framework of dielectric theory, we
will focus on the stopping power at arbitrary temperature anisotropy
$T_{\bot} /T_{\parallel}$.

The paper is organized as follows. We start in 
Sec. II, with solving the linearized Vlasov-Poisson equations by means of Fourier
transformation. This provides the general form of the linearized potential
generated in a temperature anisotropic magnetized Maxwellian plasma by a
projectile ion from which the stopping power is deduced.

In the next Sec. III, is dedicated to apply our results to nonmagnetized plasma. Calculations
are carried out for small projectile velocities at arbitrary temperature
anisotropy and arbitrary direction of ion motion with respect to
the anisotropy axis.

Then we turn to the effect of a weak magnetic field on the stopping power 
in Sec. IV, while we concentrate on the influence of a strong magnetic field 
in Sec. V. In contrast with the papers \cite{hbn,cs} we consider an ion 
motion in arbitrary direction. 

As the last issue we investigate in Sec. VI the stopping power for small
projectile velocities at arbitrary magnetic field and temperature anisotropy. 
The friction coefficient there contains an anomalous term which increases
logarithmically when the projectile velocity approaches to zero.

The achieved results are finally summarized and discussed in Sec. VII.

\section{DIELECTRIC THEORY}

For the temperature anisotropic plasma with two different temperatures
$T_{\parallel}$, $T_{\bot}$ of the electrons we define an average temperature
$\overline{T} = \frac{1}{3} T_{\parallel} + \frac{2}{3} T_{\bot}$.  Within the
dielectric theory the electron plasma is described as a continuous,
polarizable fluid (medium), which is represented by the phase-space
density of the electrons $f({\bf r}, {\bf v}, t)$. Here, 
only a mean-field interaction between the electrons is considered and
hard collisions are neglected and the evolution of the distribution function 
$f({\bf r}, {\bf v},t)$ is determined by the Vlasov-Poisson equation 
is valid for weakly coupled plasmas where the number of electrons in the Debye sphere
$N_{\rm D} = 4 \pi n_0 \overline{\lambda}_{\rm D}^3 \gg 1$ is very large. 
Here $n_0$ is the electron density, $\overline{\lambda}_{\rm D}= (k_{\rm B}
\overline{T}/ 4 \pi n_0 e^2)^{1/2}$ is an averaged Debye length.

In the following, we consider a nonrelativistic projectile ion with charge $Ze$ and
with a velocity $\bf V$ that moves in a magnetized temperature anisotropic
plasma at an angle $\vartheta$ with respect to the magnetic field
$\bf B_0$. The axis defined by $\bf B_0$ also coincides with the degree of 
freedom with temperature $T_{\parallel}$. We assume that the energy variation
of the ion is much smaller than its kinetic energy. The strength of the coupling
betweeen an ion moving with velocity $V$ and the electron plasma is given by
the coupling parameter

\begin{equation}
{\cal{Z}} = \frac{|Z|}{N_{\rm D} \left[ 1+
    V^2/\overline{v}_{\rm th}^2 \right]^{3/2}} .  
\end{equation}
Here $\overline{v}_{\rm th} = (k_{\rm B}\overline{T}/m)^{1/2}$ 
is the average thermal velocity of an electron. The derivation of Eq. (1) is discussed in
detail in Ref. \cite{gz}. The parameter ${\cal{Z}}$
characterizes the
ion-target coupling, where ${\cal{Z}} \ll 1$ corresponds to weak,
almost linear coupling and ${\cal{Z}} \gtrsim 1$ to strong,
nonlinear coupling.

For a sufficiently small perturbation (${\cal{Z}} \ll 1$) the
linearized Vlasov equation of the plasma may be written as

\begin{equation}
\frac{\partial f_1}{\partial t} + {\bf v} \ \frac{\partial f_1}{\partial {\bf r}} - \omega_c
\left[ {\bf v} \times {\bf b} \right] \frac{\partial f_1}{\partial {\bf v}} =
- \frac{e}{m} \ \frac{\partial \varphi}{\partial {\bf r}} \ \frac{\partial
  f_0}{\partial {\bf v}} ,
\end{equation}
where $f=f_0 +f_1$ and the self-consistent electrostatic potential $\varphi$ is determined by
the Poisson equation

\begin{equation}
\bigtriangledown^2 \varphi = -4 \pi Ze \delta ({\bf r}-{\bf V}t) + 4 \pi e 
\int d {\bf v} f_1 ({\bf r}, {\bf v},t) .
\end{equation}
The $\bf b$ is the unit vector parallel to ${\bf B}_0, -e$ and
$\omega_c = eB_0/mc$ are the charge and Larmor frequency of plasma
electrons respectively, $f_0$ is the unperturbed distribution function
of plasma electrons, which in the case of temperature anisotropic,
homogeneous electron plasma is given by two Maxwellians for the
longitudinal and transversal degrees of freedom

\begin{equation}
f_0 (v_{\parallel}, v_{\bot})= 
\frac{n_0}{(2 \pi)^{3/2}v^2_{{\rm th} \bot} v_{{\rm th} \parallel}} 
\exp \left( - \frac{v_{\bot}^2}{2 v^2_{{\rm th}\bot}} \right)
\exp \left(- \frac{v_{\parallel}^2}{2v_{{\rm th} \parallel}^2} \right) \ ,
\end{equation}
where $\langle v_{\parallel}^2 \rangle = v_{{\rm th} \parallel}^2 = k_{\rm B}T_{\parallel}/m,
\quad \langle v_{\bot}^2 \rangle = 2v_{{\rm th} \bot}^2 = 2k_{\rm B}T_{\bot}/m$.

By solving Eqs. (2) and (3) in space-time Fourier components, we obtain 
the electrostatic potential

\begin{equation}
\varphi ({\bf r},t)= \frac{Ze}{2 \pi^2} \int d {\bf k} \  
\frac{\exp \left[ i {\bf k} ({\bf r}-{\bf V}t)\right]}
{k^2 \varepsilon ({\bf k}, {\bf kV})} \ ,
\end{equation}
which provides the dynamical response of the temperature
anisotropic plasma to the motion of the projectile ion in the presence
of the external magnetic field. Here $\epsilon ({\bf k}, \omega)$ is the 
dielectric function of a temperature anisotropic, magnetized plasma which 
is given by

\begin{eqnarray}
\varepsilon ({\bf k}, \omega) 
&=& 1 + \frac{1}{k^2 \lambda^2_{{\rm D} \parallel}} 
\left[ G(s) +i F(s) \right] \\ \nonumber 
&=& 1 +  \frac{1}{k^2 \lambda^2_{{\rm D} \parallel}} 
\left\{ 1+ is \sqrt{2} \int\limits^{\infty}_0 dt
\exp \left[ ist \sqrt{2} - X(t) \right] \right.  \\ \nonumber
&+& \left. \frac{kv_{{\rm th} \parallel} \sqrt{2}}{\omega_c} \sin^2 \alpha (1- \tau) 
\int\limits_0^ {\infty} dt
 \sin \left( \frac{\omega_c t \sqrt{2}}{kv_{{\rm th} \parallel}} \right)
\exp \left[ ist \sqrt{2} - X(t) \right] \right\}
\end{eqnarray}
with

\begin{equation}
X(t) = t^2 \cos^2 \alpha + k^2 a^2_{c \bot} \sin^2 \alpha
\left[ 1- \cos \left( \frac{\omega_c t \sqrt{2}}{kv_{{\rm th} \parallel}}\right) \right] \ ,    
\end{equation}
where $\lambda_{{\rm D} \parallel} = v_{{\rm th}
  \parallel}/\omega_p$, $\omega_p$ is the plasma frequency, $s=
\omega/kv_{{\rm th} \parallel}$, $\tau = T_{\bot}/T_{\parallel}$,
$a_c = v_{{\rm th} \bot} / \omega_c$ and $\alpha$ is the angle between
the wave vector $ \bf k$ and the magnetic field.

As shown in Appendix A, Eqs. (6) and (7) are identical with the
Bessel function representation of $\varepsilon ({\bf k},\omega)$ derived e.g. 
by Ichimaru \cite{si}. Eqs. (6) and (7) are, however, more convenient when
studying the weak and strong magnetic field limits in Secs. IV and V.

The stopping power $S$ of an ion is defined as the energy loss of
the ion in a unit length due to interactions with the plasma
electrons. From Eq. (5) it is straightforward to calculate the electric
field ${\bf E} = - {\bf \bigtriangledown} \varphi $, and the stopping
force acting on the ion. Then, the stopping power of the projectile ion becomes
 
\begin{eqnarray}
 S &=& - \frac{dE}{dl} = \left. Ze \frac{\partial}{\partial {\bf r}}
  \varphi ({\bf r},t) \right|_{{\bf r} ={\bf V}t} \\ \nonumber
&=& \frac{2Z^2e^2 \lambda_{{\rm D} \parallel}^2}{\pi^ 2}
\int\limits_0^{k_{\rm max}} k^3 dk \int\limits_0^{1} d \mu
\int\limits_0^{\pi} d \varphi \frac{\cos \Theta F(s)}{[k^2
\lambda_{{\rm D} \parallel}^2+ G(s)]^2 + F^2(s)} \ , 
\end{eqnarray}
where $\mu = \cos \alpha$ was the angle between $\bf k$ and ${\bf B}_0$, 
$\Theta$ is the angle between $\bf k$ and $\bf V$, $s= {\bf k} \cdot {\bf V}/kv_{{\rm th} \parallel} = (V/v_{{\rm th} \parallel}) \cos\Theta$, 
$ \cos \Theta = \mu \cos \vartheta - \sqrt{1-\mu^2} \sin \vartheta \cos \varphi$, and 
$\vartheta$ is the angle between $\bf V$ and ${\bf B}_0$. 
In Eq. (8) we introduced a cutoff parameter
$k_{\rm max} = 1/r_{\rm min} $ (where $r_{\rm min} $ is the
effective minimum impact parameter) in order to avoid the logarithmic
divergence at large $k$. This divergence corresponds to the
incapability of the linearized Vlasov theory to treat close encounters
between the projectile ion and the plasma electrons properly.
For $r_{\rm min}$ we thus use the effective minimum impact parameter
of classical binary Coulomb collisions $r_{\rm min} = Ze^2/mv_{r}^2$ 
for relative velocities $v_{r} \simeq (V^2 + \overline{v}^2_{\rm th})^{1/2}$, 
which is often called the ``distance of closest approach.'' Hence

\begin{equation}
k_{\rm max} = \frac{1}{r_{\rm min}} = \frac{m(V^2 + 
\overline{v}^2_{\rm th})}{Ze^2} .
\end{equation}

A two temperature description of an electron plasma is valid only when the
ion beam-plasma interaction time
is less than the relaxation time between the two temperatures,
$T_{\parallel}$ and $T_{\bot}$. For an estimate we will briefly consider the field-free
case, because the external magnetic field suppresses the relaxation
between the transversal and longitudinal temperatures during the time
of flight of the ion beam through plasma.

The problem of a temperature relaxation in a temperature anisotropic
plasma with and without of an external magnetic field was considered
by Ichimaru \cite{si}.  Within the dominant-term approximation the relaxation
time $\Delta \tau_{\rm rel}$ for the plasma without magnetic field is given by

\begin{equation}
\frac{1}{\Delta \tau_{\rm rel}} = \frac{8}{15} \sqrt{\frac{\pi}{m}} \ 
\frac{n_0 e^4}{(k_{\rm B}T_{\rm eff})^{3/2}} \ \ln \Lambda_c \ ,
\end{equation} 
where $\ln \Lambda_c = \ln (N_D)$ is the Coulomb logarithm and
the effective electron temperature $T_{\rm eff}$ is defined through

\begin{eqnarray}
\frac{1}{T_{\rm eff}^{3/2}} = \frac{15}{2} \int\limits_0^1
\frac{\mu^2 (1- \mu^2) d \mu}{[\mu^2 T_{\parallel} + (1- \mu^2) T_{\bot}]^{3/2}} \\ \nonumber
= \frac{5 \sqrt {3}}{12 \overline {T}^{3/2}} \frac{(1+2\tau )^{3/2}}{(\tau -1)^2} 
\left[ \frac{\tau + 2}{\sqrt{|\tau -1|}} \ p_0 (\tau) -3 \right] \ , 
\end{eqnarray}

\begin{equation}
p_0 (\tau) = \left\{
\begin{array}{ll}
\ln \frac{1+ \sqrt{1- \tau}}{\sqrt{\tau}} \ , \qquad \qquad \tau < 1 \\
\arctan \sqrt{\tau -1} \ , \qquad \tau > 1
\end{array} \right. .
\end{equation}

The relaxation time calculated from Eq. (11) are of the order of
$10^{-6}$s, $0.5\times 10^{-5}$s and $10^{-3}$s for averaged temperatures 
$\overline {T} =10^{-2}$eV, $\overline {T} =0.1$eV and 
$\overline {T} =1$eV, respectively, for anisotropies $\tau \simeq 0.01-100$. 
The interaction time (for instance, for ICF or for electron cooling) is about 
$10^{-7} -10^{-8}$s. Therefore, ion beam-plasma interaction time can be very small 
compared to the plasma relaxation time.

\section{STOPPING POWER IN PLASMA WITHOUT MAGNETIC FIELD}

Let us analyse expression (8) in the case when a projectile ion moves in a temperature
anisotropic plasma without magnetic field. The plasma dielectric function from Eqs. (6)
and (7) now takes the form

\begin{equation}
\varepsilon ({\bf k},\omega) =1+ \frac{1}{k^2 \lambda^2_{{\rm D}{\parallel}}}
 \frac{1}{A^2} W \left( \frac{s}{A} \right) \ .
\end{equation}
Here $A= (\mu^2 + \tau (1- \mu^2))^{1/2}$ and $W(s) = g_0 (s) + if_0 (s)$ is the plasma
dispersion function \cite{dbf},

\begin{equation}
g_0 (s) = 1- s \sqrt{2} \ Di \left( \frac{s}{\sqrt{2}}\right); \quad
f_0 (s) = \sqrt{\frac{\pi}{2}} s \ \exp \left(- \frac{s^2}{2} \right),
\end{equation}
where

\begin{equation} 
Di (s) = \exp (-s^2) \int\limits_0^s  dt \exp (t^2)
\end{equation}
is the Dawson integral \cite{dbf} which has for large arguments $s$ the asymptotic
$Di (s) \simeq 1/2s + 1/4s^3$.

Substituting Eq. (13) into Eq. (8) and performing the $k$-integration we obtain

\begin{equation}
S_0 = \frac{Z^2e^2}{2 \pi^2 \lambda_{{\rm D} \parallel}^2} 
\int\limits_0^1  d \mu \int\limits_0^{\pi} d \varphi
\frac{\cos \Theta}{A^2} \ Q_0 \left( \frac{v}{v_{{\rm th} \parallel}}
\frac{\cos \Theta}{A} \ , \ \xi_{\parallel}A \right) \ , 
\end{equation}
where $\xi_{\parallel} = k_{\rm max} \lambda_{{\rm D} \parallel}$ and

\begin{eqnarray}
Q_0 (x, \xi) = f_0 (x) \ln
\frac{f_0^2 (x) + [\xi^2 + g_0 (x)]^2}{f_0^2 (x) + g_0^2 (x)} \\ \nonumber
+ 2g_0 (x) \left[\arctan \frac{g_0 (x)}{f_0 (x)} - \arctan \frac{\xi^2 + g_0 (x)}{f_0 (x)} \right] \ .
\end{eqnarray}

In the case of temperature isotropic plasma $(T_{\bot} = T_{\parallel} \equiv T,$ and $ \tau =1)$ $A=1$ 
and Eq. (16) coincides with the result of e.g. Ref. \cite{tp}

\begin{equation}
S_0 = \frac{Z^2 e^2}{2 \pi \lambda_D^2} \frac{v_{\rm th}^2}{V^2}
\int\limits_0^{V/v_{\rm th}} d \mu \mu Q_0 (\mu , \xi) \ , 
\end{equation} 
where $v_{\rm th} = v_{{\rm th} \parallel}= v_{{\rm th} \bot}$,
$\lambda_D = v_{\rm th}/\omega_p$, and $\xi = k_{\rm max} \lambda_D $.

When a projectile ion moves slowly through a plasma, the electrons
have much time to experience the ion attractive potential. They are
accelerated towards the ion, but when they reach its trajectory the
ion has already moved forward a little bit. Hence, we expect an
increased density of electrons at some place in the trail of the ion.
This negative charge density pulls back the positive ion and gives
rise to the stopping. This drag force is of particular interest for the
electron cooling process. In the limit of small velocities 
$S \simeq R \cdot V$. This looks like the friction law of a viscous fluid, 
and accordingly $R$ is called the friction coefficient. However, in the case
of an ideal plasma it should be noted that this law does not depend on the plasma
viscosity and is not a consequence of electron-electron collisions
which are neglected in the Vlasov equation.

The Taylor expansion of Eq. (16) for small $V (V \ll \overline {v}_{\rm th} )$ yields the friction law

\begin{equation}
S_0 = \frac{Z^2 \left( e^2/\overline {\lambda}^2_{D} \right)}{3 \sqrt{2 \pi} } 
\frac{V}{\overline {v}_{\rm th} } \psi (\overline {\xi}) 
\left[I_1 (\tau) + I_2 (\tau) \sin^2 \vartheta \right],
\end{equation}
where $\overline {\xi} =k_{\rm max} \overline {\lambda}_D =\left(1+V^2/\overline {v}_{\rm th}^2\right)/{\cal{Z}} \simeq 1/{\cal{Z}}$,

\begin{equation}
I_1 (\tau) = \frac{3}{\psi (\overline {\xi})} \left(\frac {2\tau +1}{3} \right)^{3/2} \int\limits_0^1 d \mu \ 
\frac{\mu^2  \psi (\xi_{\parallel} A (\mu))}{A^3 (\mu)} ,
\end{equation}

\begin{equation}
I_2 (\tau) =  \frac{3}{2 \psi (\overline {\xi})} \left(\frac {2\tau +1}{3} \right)^{3/2} \int\limits_0^1 d \mu \ 
\frac{(1- 3 \mu^2)  \psi (\xi_{\parallel} A (\mu))}{A^3 (\mu)} ,
\end{equation}
and the function $\psi$ is

\begin{equation}
\psi (\xi) = \ln (1+ \xi^2) - \frac{\xi^2}{1+ \xi^2}.
\end{equation}

In the case of temperature isotropic plasma $(\tau =1)$ we have $I_1=1$ and $I_2=0$. 
Then the Eq. (19) becomes the usual friction law in an isotropic
plasma \cite{tp}. For the strongly temperature anisotropic case, when 
$\tau \ll 1$ ($T_{\bot} \ll T_{\parallel}$) we have $\xi_{\parallel} \simeq \sqrt{3}/{\cal{Z}}$ and

\begin{equation}
I_1 \simeq - \frac{\sqrt{3}}{6 \psi (\overline {\xi})} 
\left[Li_2 (1+ \xi^2_{\parallel}) + \ln (1+ \xi^2_{\parallel}) \right] \ ,
\end{equation}

\begin{equation}
I_2  \simeq \frac{\sqrt{3}}{12 \psi (\overline {\xi})} 
\left[\xi^2_{\parallel} + 2 \ln (1+ \xi^2_{\parallel}) +
3 Li_2 (1+ \xi^2_{\parallel}) \right] .
\end{equation}
Here the functions $I_1$ and $I_2$ do not depend on $\tau$, and $Li_2(x)$ is the
dilogarithm function \cite{isg}. Note that ${\cal{Z}} \ll 1$ and
therefore $\overline \xi \gg 1$, $\xi_{\parallel} \gg 1$ in Eqs. (23)
and (24). The Coulomb logarithms in Eqs. (23) and (24) are then the
leading terms and 

\begin{equation}
I_1 \simeq \frac{\sqrt{3}}{6} \ln \frac{1}{\cal{Z}} \ll 
I_2 \simeq \frac{\sqrt{3}}{8{\cal{Z}}^2} \ \frac{1}{\ln (1/{\cal{Z}}) } \ .
\end{equation}

The normalized friction coefficient (Eq. (19)) is thus dominated by the second term 
and increases with increasing $\vartheta$.

In the opposite case, $\tau \gg 1$ ($T_{\bot} \gg T_{\parallel}$), 
the evoluation of Eqs. (20) and (21) yields

\begin{equation}
I_1 \simeq \frac{\pi \sqrt{6}}{3 \psi (\overline {\xi})} 
\left(\sqrt {1+\frac {3}{2} \overline {\xi}^2} -1-
2 \ln \frac{1+\sqrt {1+\frac {3}{2} \overline {\xi}^2}}{2}  \right) \ ,
\end{equation}

\begin{equation}
I_2 \simeq \frac{\pi \sqrt{6}}{6 \psi (\overline {\xi})} 
\left(1+\frac {1}{\sqrt {1+\frac {3}{2} \overline {\xi}^2}}-2
\sqrt {1+\frac {3}{2} \overline {\xi}^2} +
6 \ln \frac{1+\sqrt {1+\frac {3}{2} \overline {\xi}^2}}{2}  \right) \ ,
\end{equation}
and

\begin{equation}
I_1 \simeq -I_2 \simeq \frac{\pi}{2{\cal{Z}} \ln (1/{\cal{Z}})} .
\end{equation}
Then $I_1 +I_2 \sin^2 \vartheta \simeq I_1 \cos^2 \vartheta$ and the 
normalized friction coefficient decreases with increasing of $\vartheta$ in
this case.

In Fig. 1 the normalized friction coefficient $I_1 +I_2 \sin^2 \vartheta$ is plotted
as a function of temperature anisotropy $\tau$ for $\vartheta =0$ (solid line),
$\vartheta =\pi /6$ (dotted line), $\vartheta =\pi /3$ (dashed line),
$\vartheta =\pi /2$ (dot-dashed line) and for fixed plasma density and average
temperature (${\cal{Z}}=0.2$). Fig. 1 shows an enhancement of the friction
coefficient when the ion moves along the direction with low temperature. This 
effect can be easily explained in a binary collision picture. Let us consider the
particular case of strongly anisotropic plasma $T_{\bot} \gg T_{\parallel}$. In 
this case the plasma electrons move mostly in the direction across to the
anisotropy axis. For $\vartheta \simeq \pi/2$ the projectile ion moves along
the plasma electrons thermal fluctuation direction and effective impact parameter
for electron-ion collision is very small. Then the friction coefficient decreases.
For $\vartheta \simeq 0$ the projectile ion moves across to the direction of
plasma electrons thermal fluctuation. Therefore, the impact parameter
for electron-ion collisions increases which rises the friction coefficient.

For arbitrary projectile velocities we evaluated Eq. (16) numerically. 
In Figs. 2 and 3 the stopping power is plotted for strongly temperature anisotropic
plasmas ($\tau =10^{-2}$ and $\tau =10^2$ in Figs. (2) and (3) respectively) 
with $n_0 =10^8 {\rm cm}^{-3}$, $\overline {T} =0.1$eV and for four values
of $\vartheta$; $\vartheta =0$ (dotted line), $\vartheta =\pi/6$ (dashed line),
$\vartheta =\pi/3$ (long-dashed line), $\vartheta =\pi/2$ (dot-dashed line).
The solid lines are plotted for temperature isotropic plasma with
$T=\overline {T} =0.1$eV. The general behaviour of the stopping power for two anisotropy
parameters $\tau$ is characterized by an increase by comparision with the
isotropic case. At 
$\vartheta \simeq \pi/2$ and $\tau =10^{-2}$ (Fig. (2)) the ion moves in direction
accross to the longitudinal electron motion with the lower temperature $T_{\bot}$ and the
maximum of the stopping power is around $V \simeq v_{{\rm th} \bot}$, whereas the maximum
for an ion motion in longitudinal direction is at 
$V \simeq v_{{\rm th}\parallel} \gg v_{{\rm th}\bot}$.

\section{STOPPING IN PLASMAS WITH WEAK MAGNETIC FIELD}

For the case when the magnetic field is weak, in the sense that the
dimensionless parameter $\eta = \omega_c/ \omega_p$ is much less
than unity, the functions $G$ and $F$, Eqs. (6) and (7), which define the 
dielectric function, can be expanded about its field free values $g_0 (s/A)/A^2$, 
$f_0 (s/A)/A^2$ Eqs. (14) and (15)

\begin{equation}
G(s) + iF(s) = \frac{1}{A^2} \left[ g_0 \left(\frac{s}{A}\right)+ if_0 \left(\frac{s}{A}\right)\right]
+ \eta^2 \frac{\sin^2 \alpha}{(k \lambda_{D \parallel})^2}\left[g_1 (s) + if_1 (s) \right],
\end{equation}
where

\begin{eqnarray}
g_1 (s) + if_1 (s) &=& \frac{2}{3} (1- \tau) \int\limits_0^{\infty} t^3 dt
\left(\frac{t^2}{2} \tau \sin^2 \alpha -1 \right) \exp
(ist \sqrt{2} - A^2 t^2) \\ \nonumber
&+& \frac{is \sqrt{2}}{6} \tau \int\limits_0^{\infty} t^4 dt 
\exp (ist \sqrt{2} - A^2 t^2) ,    
\end{eqnarray}
$s= \omega/ kv_{{\rm th} \parallel}$. Substituting this expression (29)
into Eq. (8) leads to

\begin{equation}
S = S_0 + \eta^2 S_1 ,
\end{equation}
where $S_0$ is the stopping power in plasma without magnetic field Eq. (16) and 
$\eta^2 S_1$ represents the change due to a weak magnetic
field. After some simplifications it becomes

\begin{eqnarray}
S_1 &=& \sqrt{\frac{\pi}{2}}  \frac{Z^2 e^2}{24 \pi^2 \lambda^2_{D \parallel}} 
 \frac{V}{v_{{\rm th} \parallel}} \int\limits_0^1 d \mu
\int\limits_0^{\pi} d \varphi \frac{(1-\mu^2)\cos^2 \Theta}{A^5} \\ \nonumber
&\times& \exp \left(- \frac{V^2}{v_{{\rm th} \parallel}^2}   
\frac{\cos^2 \Theta}{2A^2} \right)  
\frac{\tau \left(7 - \frac{V^2}{v_{{\rm th} \parallel}^2} 
\frac{\cos^2 \Theta}{A^2}  \right) - 4A^2} 
{f_0^2 \left(\frac{V}{v_{{\rm th} \parallel}}  \frac{\cos \Theta}{A}  
\right) +
g_0^2 \left( \frac{V}{v_{{\rm th} \parallel}} 
\frac{\cos \Theta}{A}  \right) } .
\end{eqnarray}
In the temperature isotropic plasma $(\tau =1)$ Eq. (32)
coincides with the results by May and
Cramer \cite{rmm} after integration over $\varphi$. Note that the additional term $S_1$ 
does not depend on the cutoff 
parameter $k_{\rm max}$.

In the next subsections we evaluate Eq. (32) for small and large
projectile velocities.

\subsection{Small projectile velocities}

When the projectile ion moves slowly $(V< \overline {v}_{\rm th})$ in 
plasma Eq. (32) leads to the simplified expression

\begin{equation} 
S_1 = \frac{Z^2 e^2}{60 \pi \overline {\lambda}^2_{D}} \sqrt{\frac{\pi}{2}}
\frac{V}{\overline {v}_{\rm th} } P (\vartheta, \tau) ,
\end{equation}
with

\begin{equation}
P (\vartheta, \tau) = \left(\frac {1+2\tau}{3}\right)^{3/2} \left[ P_1 (\tau) + P_2 (\tau) \sin^2 \vartheta \right], 
\end{equation}

\begin{equation}
P_1 (\tau)= \frac{5}{6(1- \tau)^2} \left\{ 14 \tau + 25 -
\frac{3(9 \tau +4)}{\sqrt{|1- \tau|}} p_0 (\tau) \right\} ,
\end{equation}

\begin{equation}
P_2 (\tau)= \frac{5}{12 \tau (1- \tau)^2} \left\{ 
\frac{3 \tau (23 \tau +16)}{\sqrt{|1- \tau|}} p_0 (\tau) -28 \tau^2 -
91 \tau +2 \right\} \ .
\end{equation}
Here, the function $p_0 (\tau)$ is given by Eq. (12). In
temperature isotropic plasma with $\tau =1$ we
have $P_1 (1) = P_2 (1) = 1$.

In Fig. 4 the normalized friction coefficient $P(\vartheta,\tau)$ for the
additional stopping power $S_1$ is plotted as a function of $\tau$ for $\vartheta =0$ 
(solid line), $\vartheta =\pi/6$ (dotted line), $\vartheta =\pi/3$ 
(dashed line), $\vartheta =\pi/2$ (dot-dashed line). The general behaviour 
of $P(\vartheta, \tau)$ is similar to the friction coefficient of the 
plasma without magnetic field (see Fig. 1). Here, the correction $P(\vartheta,\tau)$
can be also negative at small $\tau$ and $\vartheta$, which then corresponds
to a slight decrease of the stopping power, Eq. (31).

\subsection{High projectile velocities}

When the projectile ion moves with large velocity ($V \gg \overline {v}_{\rm th}$), 
Eq. (32), yields

\begin{equation}
S_1 = - \frac{Z^2 e^2 \omega_p^2}{8V^2} 
\left\{ 2C_1 (1+ \cos^2 \vartheta) - C_2 B (\vartheta, \tau)
\left[ \cos^2 \vartheta + \frac{\sin^2 \vartheta}{B(\vartheta, \tau)+1} 
\right ]\right\} ,
\end{equation}
where

\begin{equation}
B(\vartheta, \tau) = \left( \frac{\tau}{\tau \cos^2 \vartheta + \sin^2 \vartheta} \right)^{1/2} ,  
\end{equation}

\begin{equation}
C_1 = \frac{1}{3 \sqrt{2 \pi}} \int\limits_0^{\infty} 
\frac{x^2 \exp (-x^2/2) dx}{f_0^2 (x) + g_0^2 (x)} \ , \ 
C_2 =  \frac{1}{3 \sqrt{2 \pi}} \int\limits_0^{\infty} 
\frac{x^2 (7-x^2) \exp (-x^2/2) dx}{f_0^2 (x) + g_0^2 (x)} .
\end{equation}
For numbers $C_1$ and $C_2$ we get the accurate values $C_1 =1$ and $C_2 =0$
($C_2 \simeq 10^{-12}$) respectively. Therefore from Eq. (37) we have finally

\begin{equation}
S_1 = - \frac{Z^2 e^2 \omega_p^2}{4V^2} (1+ \cos^2 \vartheta) .
\end{equation}
This result is in accord with the results of Honda et al. \cite{nh} and May and 
Cramer \cite{rmm}, who, however, kept the terms $O(V^{-4})$ in their work as well.
Although the function $S_1$ in Eq. (40) is proportional to
the plasma density, the full correction term $\eta^2 S_1$ does not depend
on the plasma density.

In Figs. (5) and (6) we show the velocity dependence of the function $S_1$
for $\tau =10^{-2}$ and $\tau =10^2$ respectively. The different curves are
$\vartheta =0$ (solid line), $\vartheta =\pi/6$ (dotted line), $\vartheta =\pi/3$
(dashed line), $\vartheta =\pi/2$ (dot-dashed line). For small and medium 
projectile velocities the weak magnetic
field decreases the total stopping power for small $\tau$ and increases
it in the high $\tau$ limit. For high projectile velocities 
the magnetic field always reduces the stopping power independent of the 
temperature anisotropy, see Eq. (40).

\section{STOPPING IN PLASMAS WITH STRONG MAGNETIC FIELD}

We now turn to the case when a projectile ion moves
in a temperature anisotropic plasma with a strong
magnetic field, which is on one hand, sufficiently weak to allow a 
classical description $(\hbar \omega_c < k_{\rm B}
T_{\bot}$ or $ \hbar /mv_{{\rm th}\bot} < a_c$, and, on the other hand, 
comparatively strong so that the
cyclotron frequency of the plasma electrons exceeds the plasma
frequency $\omega_c \gg \omega_p$. This limits the values of the magnetic field itself
and values of perpendicular temperature and plasma density. From these
conditions we can obtain

\begin{equation} 
3 \times 10^{-6} n_0^{1/2} < B_0 < 10^5 T_{\bot} ,   
\end{equation}
where $n_0$ is measured in cm$^{-3}$, $T_{\bot}$ is measured in eV, and
$B_0$ in kG. Conditions (41) are always true in the range of
parameters $n_0 < 10^{15}$cm$^{-3}$, $B_0 < 100$kG, $T_{\bot} >
10^{-3}$eV. Then the 
perpendicular motion of the electrons is completely quenched and 
the stopping power depends only on the longitudinal electron
temperature $T_{\parallel}$. The dependence on the transversal temperature
will be only introduced by the cutoff parameter Eq. (9).

In the limit of sufficiently strong magnetic field, Eq. (8) becomes

\begin{equation}
S_{\rm inf} = \frac{2Z^2 e^2}{\pi^2 \lambda_{D \parallel}^2} 
\int\limits_0^{\xi_{\parallel}} k^3 dk \int\limits_0^1 d \mu 
\int\limits_0^ {\pi} d \varphi \frac{\cos \Theta f_0 (s)}{\left[k^2 + g_0 (s)\right]^2 + f^2_0 (s)},
\end{equation}
with $s=(V/v_{{\rm th}\parallel})(\cos \Theta /\mu)$ and $g_0$, $f_0$ from Eqs. (14), 
which gives after integration over $k$

\begin{equation}
S_{\rm inf} = \frac{Z^2 e^2}{2 \pi^2 \lambda_{D \parallel}^2} 
\int\limits_0^1 d \mu \int\limits_0^{\pi} d \varphi
\cos \Theta Q_0 \left( \frac{V}{v_{{\rm th} \parallel}} 
\frac{\cos \Theta}{\mu}, \xi_{\parallel} \right).
\end{equation}
Here the function $Q_0$ is given by Eq. (17). For further
simplification of Eq. (43) we introduce the new variable of integration
$x= \cos \Theta/\mu$. After $\varphi$ integration in Eq. (43) we finally find 
the stopping power in the presence of a strong magnetic field as

\begin{equation}
S_{\rm inf} (V, \vartheta ) = 
\frac{Z^2 e^2}{8 \pi \lambda_{D \parallel}^2} Q \left(\frac{V}{v_{{\rm th} \parallel}}, \vartheta\right),
\end{equation}
where

\begin{equation}
Q \left(\frac{V}{v_{{\rm th} \parallel}}, \vartheta\right) = 
\sin^2 \vartheta \int\limits_{- \infty}^{\infty}
\frac{Q_0 \left(\frac{V}{v_{{\rm th} \parallel}} x, \xi_{\parallel}\right) xdx}
{(x^2 +1-2x \cos \vartheta)^{3/2}}.
\end{equation}

In the previous works \cite{hbn,cd,cs} only the case of
$\vartheta =0$ the motion of the projectile along the magnetic field
direction has been investigated. In this case the integral
in Eq. (45) diverges, while prefactor $\sin^2 \vartheta$ 
tends to zero. Introducing the new variable of the integration in Eq. (45)
$y=(x- \cos \vartheta)/\sin \vartheta$ we obtain for vanishing angle $\vartheta$

\begin{equation}
Q\left(\frac{V}{v_{{\rm th} \parallel}}, \vartheta \rightarrow 0 \right) =
2 Q_0 \left(\frac{V}{v_{{\rm th} \parallel}}, \xi_{\parallel} \right).
\end{equation}

Thus expression (44) reproduces the known results for the stopping power on an 
ion which moves along the direction of the magnetic field \cite{hbn,cd,cs}.

In the following paragraphs we will discuss its low and high velocity limits.

\subsection{Small projectile velocities}

In the low velocity limit ($V \ll v_{{\rm th} \parallel}$) Eq. (45) becomes

\begin{equation}
Q\left(\frac{V}{v_{{\rm th} \parallel}}, \vartheta \right) \simeq \frac{2V}{v_{{\rm th} \parallel}}
\left\{ \sqrt{2 \pi} \psi (\xi_{\parallel})
\left[ \sin^2 \vartheta \ln
\left( \frac{2 v_{{\rm th} \parallel}}{V\sin\vartheta} \right) +1
-2 \sin^2 \vartheta \right] 
+  C_1 (\xi_{\parallel}) \sin^2 \vartheta \right\},
\end{equation}
where

\begin{equation}
C_1 (\xi_{\parallel}) = \int\limits_0^1 \frac{dx}{x^2}
\left[ Q_0 (x, \xi_{\parallel}) - \sqrt{2 \pi} \psi (\xi_{\parallel}) x \right]
+ \int\limits_1^{\infty} \frac{dx}{x^2} Q_0 (x, \xi_{\parallel}) . 
\end{equation}
Here, the function $\psi $ is defined by Eq. (22). Since we deal with 
small ion beam-plasma coupling ${\cal{Z}} \ll 1$ we have, $\xi_{\parallel} \gg 1$ 
in Eqs. (47) and (48) and the function $C_1 (\xi)$ simplifies

\begin{equation}
C_1 (\xi_{\parallel}) \simeq \sqrt{2 \pi} \ln \frac{2}{\gamma} \ln \xi_{\parallel} + 0.6 ,
\end{equation}
where $\gamma =0.5772$ is Euler's constant.

We note that the friction coefficient $S_{\rm inf} /V$ from Eqs. (44) and (47) contains 
a logarithmically large term which vanishes for $\vartheta \rightarrow 0$. It will be shown in
the next section that this behaviour is a characteristic feature of the stopping power 
at low velocities and the friction coefficient for arbitrary strength of the magnetic field.

\subsection{High projectile velocities}

In the case of high projectile velocities $(V \gg v_{{\rm th} \parallel} )$ the general 
expression (45) becomes

\begin{equation}
Q \left(\frac{V}{v_{{\rm th} \parallel}}, \vartheta \right) \simeq 
\frac{4 \pi v^2_{{\rm th} \parallel}}{V^2} 
\left\{ \sin^2 \vartheta
\left[  \ln \left( \frac{2 V}{v_{{\rm th} \parallel} \sin \vartheta} \right) +
C_2 (\xi_{\parallel}) -2 \right] +1  \right\} ,
\end{equation}
where

\begin{equation}
C_2 (\xi_{\parallel})  = \frac{1}{2 \pi} \int\limits_0^1 
Q_0 (x, \xi_{\parallel}) xdx + \int\limits_1^{\infty} \frac{dx}{x} 
\left[ \frac{x^2}{2 \pi} Q_0 (x, \xi_{\parallel}) -1 \right] 
\end{equation}
which gives for $\xi_{\parallel} \gg 1$ $C_2 (\xi_{\parallel}) \simeq \ln \xi_{\parallel}$. 
The stopping power for strong magnetic fields shows in the low and high velocity limits (Eqs. (47) and
(50)) an enhancement for ions moving transversal to the magnetic field compared to the case of the
longitudinal motion ($\vartheta =0$). This effect is in agreement with
PIC simulation results \cite{mw}. In contrast to the field-free case, at strong magnetic field and
for $\vartheta =0$, $V \gg v_{\rm th{\parallel}}$ (Eqs. (44) and (50)) we have 
$S_{\rm inf} \simeq Z^2 e^2 \omega_{p}^2 /2V^2$ independent of $k_{\rm max}$. The 
cutoff $k_{\rm max}$ necessary at low ion velocities
is, however, less well defined here than for the field-free case, where
the cutoff (9) was deduced from the binary collision picture. Now,
the electrons are forced to move parallel to ${\bf B}_0$. Since we
assumed the motion of the ion in this direction as well the ion and
an electron just pass each other along a straight line. For symmetry
reasons the total momentum transfer and the stopping power is zero. Purely
binary interactions contribute nothing and the stopping of the ion
is only due to the collective response of the plasma, that is, due
to modes with long wavelengths $k<1/\lambda_{D\parallel}$. This
suggests taking $k_{\rm max}$ of the order of $1/\lambda_{D\parallel}$,
but further investigations are clearly needed here for a more precise
description in this particular case.

In Figs. (7) and (8), the stopping power $S_{\rm inf}$ is plotted as a
function of projectile velocity (in units of $v_{{\rm th} \parallel}$)
for $n_0 = 10^6$cm$^{-3}$, $T_{\parallel} = 10^{-4}$eV , $T_{\bot} =
10^{-5}$eV (Fig. (7)), $T_{\bot} = 0.1$eV (Fig. (8)), and for four
different values of angle $\vartheta: \vartheta = 0$ (solid line), $
\vartheta = \pi/6$ (dotted line), $\vartheta = \pi/3$ (dashed line)
and $\vartheta = \pi/2$ (dash-dotted line). The enhancement of $S_{\rm inf} (V,\vartheta)$ 
with respect to $S_{\rm inf} (V,0)$ in the low and in high velocity limit by increasing of the angle
$\vartheta$ is documented in Fig. (9), for $T_{\parallel}= 10^{-4}$eV, $T_{\bot} =
0.1$eV, $n_0= 10^6$cm$^{-3}$, $\vartheta = \pi/6$ (solid line),
$\vartheta = \pi/4$ (dotted line), $\vartheta = \pi/3$ (dashed line) and
$\vartheta = \pi/2$ (dash-dotted line). The physical origin of this 
angular behaviour in the low and high velocity limits is the enhancement
of the effective impact parameter for an individual electron-ion collision
with increasing $\vartheta$. For medium projectile velocities
$V \simeq v_{\rm th\parallel}$ the collective excitations in plasma
become important and then stopping power is higher for small $\vartheta$.

\section{STOPPING AT ARBITRARY MAGNETIC FIELD AND IN LOW-VELOCITY
LIMIT. ANOMALOUS FRICTION COEFFICIENT}

We now proceed with a projectile ion at low velocities and at arbitrary magnetic field. 
This regime is of particular importance for the electron 
cooling process \cite{ahs,hp,inm}. In the presence of a magnetic field the friction
coefficient here contains a term which diverges like $\ln (v_{\rm th \parallel} /V)$ in
addition to the usual (see e.g. Sec. III) constant one.

For this consideration it is convenient to use the Bessel function representation 
of the dielectric function which has been given e.g. by Ichimaru \cite{si}, 
see Appendix A Eq. (A7), and to write the real and imaginary parts 
of Eq. (A7) separately

\begin{eqnarray}
G &=&  1- \frac{\sqrt{2} \omega}{|k_{\parallel}| v_{{\rm th} \parallel}}
\Lambda_0 (z) Di \left( \frac{\omega}{|k_{\parallel}| v_{{\rm th} \parallel} 
\sqrt{2}} \right) \\ \nonumber
&-& \frac{\sqrt{2}}{|k_{\parallel}| v_{{\rm th} \parallel}}
\sum\limits_{n=1}^{\infty}  \Lambda_n (z)
\left\{ \omega \left[ Di \left( \frac{\omega + n \omega_c}
{|k_{\parallel}| v_{{\rm th} \parallel}\sqrt{2}} \right) + 
 Di \left( \frac{\omega - n \omega_c}
{|k_{\parallel}| v_{{\rm th} \parallel}\sqrt{2}} \right) \right] \right.\\ \nonumber
&+& \left. n \omega_c \left( \frac{1}{\tau} -1 \right)
\left[ Di \left( \frac{\omega - n \omega_c}
{|k_{\parallel}| v_{{\rm th} \parallel} \sqrt{2}}  \right) - Di \left( 
\frac{\omega + n \omega_c}{|k_{\parallel}| v_{{\rm th} \parallel} \sqrt{2}}  
\right) \right] \right\} \ ,
\end{eqnarray}

\begin{eqnarray}
F = \sqrt{\frac{\pi}{2}} \left\{ \frac{\omega}
{|k_{\parallel}| v_{{\rm th} \parallel}} \Lambda_0 (z) \exp \left( - 
\frac{\omega^2}{2k_{\parallel}^2 v_{{\rm th} \parallel}^2} \right) \right. 
\\ \nonumber
+ \frac{2}{|k_{\parallel}| v_{{\rm th} \parallel}} 
\sum\limits_{n=1}^{\infty} \Lambda_n (z) \exp
\left( - \frac{\omega^2+n^2 \omega_c^2}
{2k^2_{\parallel} v^2_{{\rm th} \parallel}} \right) \\ \nonumber
\left. \times \left[ \omega {\rm ch} \left( \frac{n \omega_c \omega}
{k^2_{\parallel} v^2_{{\rm th} \parallel}} \right) + n \omega_c
\left( \frac{1}{\tau} -1 \right) {\rm sh} 
\left( \frac{n \omega_c \omega}{k^2_{\parallel} v^2_{{\rm th} \parallel}}
\right) \right] \right\} \ . 
\end{eqnarray}
The notations in Eqs. (52) and (53) are explained in Appendix A. 

For the friction coefficient we have to consider $S$, given by Eq. (8)
in the low-velocity limit and thus the functions $G$ and $F$
given by Eqs. (52) and (53), when $\omega = {\bf kV}$.
Now we have to write the Taylor
expansion of Eqs. (52) and (53) for small $\omega = {\bf kV}$.
However, the first term of Eq. (53) exhibits a singular behaviour
in the limit of $\omega = {\bf kV} \rightarrow 0$ where the 
$k_{\parallel}$ integration diverges logarithmically for small
$k_{\parallel}$. We must therefore keep $\omega = {\bf kV}$
finite in that integration to avoid such a divergence. This anomalous 
contribution which arises from the first term of Eq. (53) in
low-velocity limit is

\begin{equation} 
S_{\rm an} \simeq \left( \frac{2}{\pi^3} \right)^{1/2} 
\frac{Z^2 e^2}{\lambda_{D \parallel}^2}  
\frac{V}{v_{{\rm th} \parallel}}  \int\limits_0^{\xi_{\parallel}} k^3dk
\int\limits_0^1 \frac{d \mu}{\mu} \int\limits_0^{\pi} d \varphi \cos^2 \Theta
\frac{\Lambda_0 (z) 
\exp \left(- \frac{V^2}{2v^2_{{\rm th}\parallel}} 
\frac{\cos^2 \Theta}{\mu^2} \right)}{[k^2 + E_2 (k, \mu)]^2} \ ,     
\end{equation}
where $\Lambda_0 (z) = \exp (-z) I_0 (z)$ and $E_2 (k, \mu) =
G(\omega=0)$ is

\begin{equation}
E_2 (k, \mu) = 1+ \frac{2 \sqrt{2} \eta}{k \mu} \left(\frac{1}{\tau}-1 \right)
\sum\limits_{n=1}^{\infty} n \Lambda_n (z) Di \left( \frac{n \eta}{k \mu \sqrt{2}}\right) \ . 
\end{equation}
Here $z= (k^2 \tau/\eta^2)(1-\mu^2)$, $\mu = \cos \alpha = k_{\parallel}/k$, and 
$\Theta$ is the angle between $\bf k$ and $\bf V$. After $\mu$ and $\varphi$ integration, 
see Appendix B, Eq. (54) reads

\begin{equation}
S_{\rm an} \simeq \left( \frac{2}{\pi} \right)^{1/2}
\frac{Z^2 e^2}{4 \lambda_{D \parallel}^2}  
\frac{V}{v_{{\rm th} \parallel}} \sin^2 \vartheta \ln
\left( \frac{v_{{\rm th} \parallel}}{V}  
\frac{2.26}{\sin \vartheta} \right) {\cal{F}} (\tau, \eta) ,
\end{equation}
with

\begin{equation}
{\cal{F}} (\tau, \eta) = \int\limits_0^{\tau \xi_{\parallel}^2} 
\frac{ \Lambda_0 (x/ \eta^2) xdx}{[x+1+ (\tau -1) \Lambda_0 (x/\eta^ 2)]^2} . 
\end{equation}
The function $\cal{F}$ and thus $S_{\rm an}$ (56) vanishes in the limit $B_0
\rightarrow 0$ (or $\eta \rightarrow 0$) like

\begin{equation}
{\cal{F}} (\tau, \eta) \simeq \frac{\eta}{(2 \pi)^{1/2}} 
\left[ \arctan (k_{\rm max} \lambda_{D \bot}) - 
\frac{k_{\rm max} \lambda_{D \bot}}{1 + (k_{\rm max} \lambda_{D \bot})^2}\right] \ .
\end{equation}

The anomalous term Eqs. (56) and (57) therefore represents a new effect arising from the
presence of the magnetic field, which is not restricted to anisotropic plasmas.

For temperature isotropic plasma $(\tau = 1)$ and for a 
sufficiently weak magnetic field $\eta < \xi_{\parallel}$ (or 
$\omega_c < k_{\rm max} v_{{\rm th} \parallel}$), Eq. (57) takes the form 

\begin{equation}
{\cal{F}} (\tau, \eta) \simeq \exp \left( \frac{1}{\eta^2} \right)
\left[ \left( 1+ \frac{1}{\eta^2} \right) K_0 \left( \frac{1}{\eta^2} \right)
- \frac{1}{\eta^2} K_1 \left(\frac{1}{\eta^2} \right)\right] \ , 
\end{equation}
where $K_0$ and $K_1$ are the modified Bessel functions of the second kind. In the 
case of very strong magnetic field $\eta > \xi_{\parallel} \sqrt{\tau}$ 
(or $\omega_c > k_{\rm max} \lambda_{D \bot}$) the function ${\cal{F}} (\tau, \eta)$ reads

\begin{equation} 
{\cal{F}} (\tau, \eta) \simeq \Psi (\xi_{\parallel}) = \ln (1+ \xi^2_{\parallel}) -
\frac{\xi^2_{\parallel}}{1+ \xi^2_{\parallel}} \ .
\end{equation}

The physical origin of such an anomalous friction coefficient may be
traced to the spiral motion of the electrons along the magnetic field lines. 
These electrons naturally tend to couple strongly with
long-wavelength fluctuations (i.e., small $k_{\parallel}$) along the
magnetic field. In addition, when such fluctuations are characterized
by slow variation in time (i.e., small $\omega = {\bf kV}$),
the contact time or the rate of energy exchange between the electrons
and the fluctuations will be further enhanced. In a plasma such
low-frequency fluctuations are provided by the slow projectile
ion. The above coupling can therefore be an efficient mechanism of
energy exchange between the electrons and the projectile ion. In the
limit of $V \rightarrow 0$, the frequency $\omega = {\bf kV}
\rightarrow 0$ tends to zero as well. The contact time thus becomes infinite and the
friction coefficient diverges.

The anomalous friction coefficient (see 
Eq. (56)) vanishes, however, when the ion moves along the magnetic field
($\vartheta = 0$). Then the friction coefficient is solely given by the 
second term of Eq. (53). The contribution of this term to the stopping power 
leads to the usual friction law in plasma and reads for arbitrary angles $\vartheta$

\begin{eqnarray}
S \simeq \left( \frac{2}{\pi} \right)^{1/2} 
\frac{2 Z^2 e^2}{\lambda_{D \parallel}^2} 
\frac{V}{v_{{\rm th} \parallel}}  \int\limits_0^{\xi_{\parallel}} k^3dk
\int\limits_0^1 \frac {d\mu}{\mu} 
\frac{E_1 (k, \mu)}{[k^2 + E_2 (k, \mu)]^2} \\ \nonumber
\times \left[\mu^2 \cos^2 \vartheta + \frac{1}{2} \left(1-\mu^2 \right) \sin^2 \vartheta \right] \    
\end{eqnarray}
with

\begin{equation}
E_1 (k, \mu) = \sum\limits_{n=1}^{\infty} \Lambda_n (z)
\exp \left(- \frac{n^2 \eta^2}{2k^2 \mu^2} \right)
\left[1+ \left(\frac{1}{\tau} -1 \right) \frac{n^2 \eta^2}{k^2 \mu^2} \right]
\end{equation}
and $E_2 (k, \mu)$ as defined by Eq. (55).

In Figs. (10) and (11) we compare the anomalous term $S_{\rm an}$ with the low velocity stopping 
without magnetic field $S_0$ see Eq. (19), where $S_{\rm an} /S_0$ is plotted as a function of
$\omega_c /\omega_p$ for $\vartheta =\pi/6$ (solid line), $\vartheta =\pi/3$
(dotted line), $\vartheta =\pi/2$ (dashed line), ${\cal{Z}}=0.1$, 
$V/\overline {v}_{\rm th} =0.2$, and for two values of the anisotropy
parameter $\tau$: $\tau =0.1$ (Fig. (10)), $\tau =10$ (Fig. (11)). 

We conclude that the anomalous term $S_{\rm an}$ gives espessially for strong 
magnetic fields ($\omega_c > \omega_p$) and for strongly temperature anisotropic
plasma ($T_{\bot} \gg T_{\parallel}$) an important contribution to the stopping. 
It should be noted that the observed enhancement of stopping due to $S_{\rm an}$ 
for $T_{\bot} \gg T_{\parallel}$ can be potentially interesting for future electron 
cooling experiments.

\section{SUMMARY}

The purpose of this work was to analyze the stopping power of an ion
in temperature anisotropic magnetized classical plasma. A general expression
obtained for stopping power was analyzed in four particular cases: in a 
plasma without magnetic field; in a plasma with weak and very strong magnetic
fields; and in a plasma with arbitrary magnetic field and for low-velocity
projectile.

From the results obtained in Secs. III-V, we found that the stopping
power essentially depends on the plasma temperature anisotropy. In field-free
case and for small ion velocities the effect of the anisotropy results in an 
enhancement of the stopping power when the ion moves in the direction with 
low temperature.

For small projectile velocities a weak magnetic field slightly
decreases the field-free stopping power for small $\tau$, in the 
opposite case (large $\tau$) the field-free stopping power slightly increases. 
In the high-velocity limit correction to the field-free 
stopping power for weak magnetic fields is always negative and the stopping power is 
reduced by the magnetic field.

In the case of strong magnetic fields we demonstrated an enhancement of the stopping
power with increasing of $\vartheta$ for low and high-velocity regions compared to 
the case of an ion which moves along ${\bf B}_0$. 

In low-velocity limit but for arbitrary magnetic field, we find an enhanced 
stopping power compared to the field-free value mainly due to the strong coupling 
between the spiral motion
of the electrons and the long-wavelength, low-frequency fluctuations excited
by the projectile ion. This anomalous stopping power increases with the 
angle $\vartheta$ (the angle between ion velocity $\bf V$ and magnetic field
${\bf B}_0$) and depends strongly on the temperature anisotropy 
$\tau =T_{\bot} /T_{\parallel}$, as seen in Figs. (10) and (11). Although the
nature of the anomalous stopping power is conditioned by the external magnetic
field the temperature anisotropy of the plasma can intensify this effect when
$T_{\bot} \gg T_{\parallel}$ (see Fig. (11)).

This emphasizes the importance of the special role
of fluctuations with small $k_{\parallel}$ and small $\omega$ 
(small projectile velocity $V$) and as another significant contribution to the 
energy exchange processes
arising from the collective modes of plasma. Potentially, the electron 
plasma waves and the ion acoustic waves in a magnetized plasma might 
provide a significant energy-exchange mechanism between projectile ion
and plasma particles. This fact then makes it necessary to consider the
influence of plasma collective modes to anomalous stopping process.
This problem will be treated in a subsequent work.

\acknowledgments

Finally, it is pleasure to thank Prof. Christian Toepffer for 
helpful discussions. 
One of the authors (H.B.N.) is grateful to Prof.~Christian Toepffer
for hospitality at the Institut f\"ur Theoretische Physik II,
Universit\"at Erlangen-N\"urnberg, where this work was concluded and
would like to thank the Deutscher Akademischer Austauschdienst for
financial support. We are indebted to Claudia Schlechte for her help
in preparing the manuscript.

\begin{appendix}

\section{}

Here we describe the evaluation of the
dielectric function in the temperature anisotropic case where the velocity
distribution of the unperturbated distribution function was given by
Eq. (4). We next introduce the Fourier transformations of $f_1
({\bf r}, {\bf v}, t)$ with respect to variables ${\bf r}$ and
$t$, $f_1 ({\bf k}, \omega , {\bf v})$. Because of the cylindrical
symmetry (around the magnetic field direction ${\bf b} =
{\bf B}_0/B_0= \hat{\bf z}$) of the problem, we choose

\begin{equation}
{\bf v} = v_{\bot} \cos \sigma \hat{\bf x} + v_{\bot} \sin \sigma
  \hat{\bf y} + v_{\parallel}  \hat{\bf z}.
\end{equation}
Then the Vlasov Eq. (2) for the distribution function becomes

\begin{equation}
\frac{\partial}{\partial \sigma} f_1 ({\bf k}, \omega, {\bf v}) +
\frac{i}{\omega_c} ( {\bf kv} - \omega - i {\rm 0})
f_1 ({\bf k} , \omega , {\bf v}) 
 = - \frac{ie}{m \omega_c} \varphi ({\bf k} , \omega) 
\left({\bf k} \frac{\partial f_0}{\partial {\bf v}} \right),
\end{equation}
where $\varphi ({\bf k} , \omega)$ is the Fourier transformation of
$\varphi ({\bf r}, t)$. The positive infinitesimal $+i0$ in
Eq. (A2) serves to assure the adiabatic turning on of the disturbance
and thereby to guarantee the causality of the response. The solution
of the Eq. (A2) has the form

\begin{equation}
 f_1 ({\bf k}, \omega , {\bf v})  = 
 - \frac{ie}{m \omega_c} \varphi ({\bf k} , \omega) 
\int\limits^{\sigma}_{\infty} d \sigma_2
\left( {\bf k} 
\frac{\partial f_0}{\partial {\bf v}} \right)_{\sigma = \sigma_2} 
 \exp \left[ \frac{i}{\omega_c} \int\limits_{\sigma}^{\sigma_2} d \sigma_1
\left[ - \omega - i0 + ({\bf kv})_{\sigma = \sigma_1} \right] \right].
\end{equation}
Combining Eq. (A3) with the Poisson equation (3) we find for the dielectric function

\begin{equation}
\varepsilon ({\bf k}, \omega) = 1 - \frac{4 \pi ie^2}{m \omega_c k^2}
\int\limits_0^{\infty}  v_{\bot} d v_{\bot} \int\limits_0^{2 \pi} d \sigma
\int\limits^{+ \infty}_{- \infty} d v_{\parallel} 
\int\limits_{\infty}^{\sigma} d \sigma_2 \left[ k_{\parallel} \frac{\partial f_0}{\partial v_{\parallel}} +
k_{\bot} \cos (\varphi - \sigma_2) \frac{\partial f_0}{\partial v_{\bot}}
 \right] $$
$$ \times \exp
\left[\frac{i}{\omega_c} \int\limits_{\sigma}^{\sigma_2} d \sigma_1
\left[ k_{\parallel} v_{\parallel} - \omega - i0 + k_{\bot} v_{\bot} \cos 
(\varphi - \sigma_1) \right]   \right],
\end{equation}
where $k_x = k_{\bot} \cos \varphi$, $k_y = k_{\bot} \sin \varphi$.
After integration by the variables $\sigma_1$, $\sigma_2$ and $\sigma$, and
using the expression \cite{isg}

\begin{equation}
\exp (-i z \sin \theta) = \sum\limits_{n= - \infty}^{+ \infty} J_n (z)
\exp (-in \theta),
\end{equation}
where $J_n$ is the Bessel function of the $n$th order, we obtain the expression \cite{si}

\begin{equation}
\varepsilon ({\bf k} , \omega) = 1- \frac{8 \pi^2 e^2}{mk^2} 
\sum\limits_{n= - \infty}^{+ \infty} \int\limits_0^{\infty} v_{\bot} 
d v_{\bot} \int\limits_{- \infty}^{+ \infty} d v_{\parallel}
\left(\frac{n \omega_c}{v_{\bot}} \frac{\partial f_0}{\partial v_{\bot}}
+ k_{\parallel} \frac{\partial f_0}{\partial v_{\parallel}} \right) 
 \frac{J_n^2 (k_{\bot} v_{\bot}/ \omega_c)}
{n \omega_c + k_{\parallel} v_{\parallel} - \omega -i0}.
\end{equation}

Substituting Eq. (4) for the unperturbed distribution function $f_0$
into Eq. (A6) we finally results in

\begin{equation}
\varepsilon ({\bf k} , \omega) = 1 + \frac{1}{k^2 \lambda^2_{D \parallel}}
\left\{ 1 + \sum\limits_{n= - \infty}^ {+ \infty} 
\left( 1+ \frac{T_{\parallel}}{T_{\bot}} 
\frac{n \omega_c}{\omega - n \omega_c} \right)
\left[ W \left( \frac{\omega - n \omega_c}
{|k_{\parallel}| v_{{\rm th} \parallel}}\right) -1 \right] 
\Lambda_n (\beta) \right\},
\end{equation}
where $\beta = k_{\bot}^2 v_{{\rm th} \bot}^2/\omega^2_c = 
k_{\bot}^2 a^2_c $, $\Lambda_n (z) = \exp (-z) I_n (z)$, $I_n (z)$
is the modified Bessel function of the $n$th order, and $W(z)$ is the plasma
dispersion function \cite{dbf}.

To show the identity of the two forms (Eqs. (6) and (A7)) of the
dielectric function we will use the expansion in modified Bessel functions \cite{isg}

\begin{equation}
\exp (z \cos \theta) = \sum\limits_{n= - \infty}^{\infty} I_n (z)
\exp (in \theta).
\end{equation}
This allows to rewrite $\exp [-X(t)]$ with $X(t)$ from Eq. (7) as

\begin{equation}
\exp [-X(t)] = \exp (-t^2 \cos^2 \alpha)
\sum\limits_{n= - \infty}^{+ \infty} \Lambda_n (\beta)
\exp \left(\frac{in \omega_c t \sqrt{2}}{kv_{{\rm th} \parallel}} \right).
\end{equation}

Substituting Eq. (A9) into expression (6) and integration over the 
variable $t$ leads to Eq. (A7).

\section{}

We now give a more detail derivation of the anomalous term $S_{\rm an}$ 
(Eq. (56)). We start with the expression (see Eq. (54))

\begin{equation}
Q(k,\varphi , \lambda) = \int\limits_0^1 \frac{d \mu}{\mu} \Phi (\mu, k,\varphi)
 \exp \left(-\frac{\lambda^2 \phi^2 (\mu, \varphi)}{2\mu^2}\right),
\end{equation}
where $\phi (\mu, \varphi)=\cos \Theta$, $\lambda =V/v_{\rm th\parallel}$,

\begin{equation}
\Phi (\mu, k,\varphi) = \frac {\Lambda_0 (z) \cos^2 \Theta}{\left[k^2+E_2 (k,\mu)\right]^2}.
\end{equation}

For $\lambda \rightarrow 0$ a leading-term approximation of (B1) leads to

\begin{equation}
Q(k,\varphi , \lambda) \simeq \Phi (0, k,\varphi) \ln \frac {\sqrt {2}}{\lambda |\phi (0, \varphi)| 
\sqrt {\gamma}} + \rm O (1),
\end{equation}
where $\gamma$ is the Euler's constant, $|\phi (0, \varphi)|=\sin \vartheta |\cos \varphi|$,

\begin{equation}
\Phi (0, k,\varphi) = \frac {\Lambda_0 (k^2 \tau / \eta^2) \sin^2 \vartheta \cos^2 \varphi}
{\left[k^2+E_2 (k,0)\right]^2}
\end{equation}
and
\begin{equation}
E_2 (k, 0) = 1+ 2 \left(\frac{1}{\tau}-1 \right)
\sum\limits_{n=1}^{\infty} \Lambda_n (k^2 \tau / \eta^2).
\end{equation}

Using the relation \cite{si,isg}

\begin{equation}
\sum\limits_{n=-\infty}^{+\infty} \Lambda_n (z) = 1 , 
\end{equation}
the function $E_2 (k,0)$ we finally takes the form

\begin{equation}
E_2 (k,0)= \frac {1}{\tau} + \left(1 - \frac{1}{\tau} \right) \Lambda_0 (k^2 \tau / \eta^2) . 
\end{equation}

Substituting Eqs. (B3), (B4) and (B7) into Eq. (54) and integration over $\varphi$ 
we finally come to expression (56).


\end{appendix}

\newpage

\begin{figure}
\caption{Normalized friction coefficient $I_1 +I_2 \sin^2 \vartheta$
    (see Eqs. (19)-(21)) in plasma with ${\cal{Z}} = 0.2$ as
    a function of $\tau = T_{\bot} / T_{\parallel}$ for
    four values of $\vartheta; \vartheta =0$ (solid line),
    $\vartheta = \pi/6$ (dotted line), $\vartheta = \pi/3$
    (dashed line), $\vartheta= \pi/2$ (dot-dashed line).}
\label{fig1}
\end{figure}

\begin{figure}
\caption{Stopping power (in units of $10^{-3}$eV/cm) as a function
    of projectile velocity $V$ (in units of $\langle v_{\rm th} \rangle =\overline{v}_{\rm th}$)
    in a strongly temperature anisotropic plasma without magnetic field
    ($\overline{T}=0.1$eV, $n_0 =10^8 \rm cm^{-3}$, $\tau = 10^{-2}$)
    for four values of angle 
    $\vartheta , \vartheta =0$ (dotted line), $\vartheta = \pi/6$ 
    (dashed line), $\vartheta = \pi/3$ (long-dashed line), 
    $\vartheta = \pi/2$
    (dot-dashed line). Solid line isotropic plasma with 
    temperature
    $T=\overline{T} = 0.1$eV.}
\label{fig2}
\end{figure}

\begin{figure}
\caption{As Fig. 2, but here $\tau = 10^2$.}
\label{fig3}
\end{figure}

\begin{figure}
\caption{The function $P(\vartheta , \tau)$ (see Eqs. (33)-(36)) as a 
  function of
  $\tau = T_{\bot} / T_{\parallel}$ for four values of 
  $\vartheta ,
  \vartheta =0$ (solid line), $\vartheta = \pi/6$ 
  (dotted line),
  $\vartheta = \pi/3$ (dashed line), $\vartheta= \pi/2$ 
  (dot-dashed line).}
\label{fig4}
\end{figure}

\begin{figure}
\caption{Additional stopping power $S_1$ (in $10^{-5}$eV/cm) in
  plasma ($n_0 =10^8\rm cm^{-3}$, $\overline{T} =0.1$eV, $\tau =10^{-2}$)
  with weak
  magnetic field (see Eq. (32)) as a function of projectile 
  velocity
  $V$ (in units of $\langle v_{\rm th} \rangle = \overline{v}_{\rm th}$) for 
  $\vartheta =0$ (solid
  line), $\vartheta = \pi/6$ (dotted line), 
  $\vartheta = \pi/3$
  (dashed line), $\vartheta= \pi/2$ (dot-dashed line).}
\label{fig5}
\end{figure}

\begin{figure}
\caption{ As Fig. 5, but here $\tau =10^2$.}
\label{fig6}
\end{figure}

\begin{figure}
\caption{Stopping power $S_{\rm inf}$ (in $10^{-3}$eV/cm) in plasma ($n_0 =10^6 \rm cm^{-3}$,
  $T_{\parallel} =10^{-4}$eV, $\tau =0.1$) with strong magnetic
  field as a
  function of projectile velocity $V$ 
  (in units of $v_{{\rm th}
    \parallel}$) for $\vartheta =0$ (solid line),
  $\vartheta = \pi/6$ (dotted line), $\vartheta = \pi/3$ 
  (dashed line), $\vartheta= \pi/2$ (dot-dashed line).}
\label{fig7}
\end{figure}

\begin{figure}
\caption{ As Fig. 7, but here $\tau = 10^3$.}
\label{fig8}
\end{figure}

\begin{figure}
\caption{The ratio $ S_{\rm inf} 
  (V,\vartheta)/S_{\rm inf} (V,0)$
  as a function of projectile velocity $V$ 
  (in units of $v_{{\rm th} \parallel})$ for $T_{\parallel} 
  = 10^{-4}$eV, 
  $\tau = 10^3,
  \vartheta = \pi/6$ (solid line), $\vartheta = \pi/4$ 
  (dotted line),
  $\vartheta= \pi/3$ (dashed line), $\vartheta= \pi/2$ 
  (dot-dashed line).}
\label{fig9}
\end{figure}

\begin{figure}
\caption{  The ratio of the anomalous stopping power to the stopping power without magnetic field
  ($S_{an} /S_0$) as a function of $\omega_c /\omega_p$ for ${\cal{Z}}=0.1$,
  $V/\overline {v}_{\rm th} =0.2$, $\tau =0.1$, $\vartheta = \pi/6$ (solid line), 
  $\vartheta = \pi/3$ (dotted line), $\vartheta= \pi/2$ (dashed line).}
\label{fig10}
\end{figure}

\begin{figure}
\caption{ As Fig. 10, but here $\tau =10$.}
\label{fig11}
\end{figure}

\end{document}